\documentclass[aps, prl, twocolumn, showpacs, nofootinbib,superscriptaddress]{revtex4}
\usepackage{graphicx}
\usepackage{epsfig}
\usepackage{amsmath}
\usepackage{amsfonts}
\usepackage{amssymb}
\usepackage{hyperref}

\newcommand{\ket}[1]{|{#1}\rangle}

\newcommand{\beq}{\begin{equation}}
\newcommand{\eeq}{\end{equation}}
\newcommand{\bea}{\begin{eqnarray}}
\newcommand{\eea}{\end{eqnarray}}
\newcommand{\tr}{\operatorname{Tr}}
\newcommand{\re}{\operatorname{Re}}
\newcommand{\im}{\operatorname{Im}}

\begin{document}

\author{Esteban A. Martinez} 
\affiliation{Departamento de F\'\i sica, FCEyN, UBA, 
Ciudad Universitaria Pabell\'on 1, 1428 Buenos Aires, Argentina} 
\author{Juan Pablo Paz} 
\affiliation{Departamento de F\'\i sica, FCEyN, UBA, 
Ciudad Universitaria Pabell\'on 1, 1428 Buenos Aires, Argentina} 
\affiliation{IFIBA CONICET, UBA, FCEyN, UBA, 
Ciudad Universitaria Pabell\'on 1, 1428 Buenos Aires, Argentina} 

\title{Dynamics and thermodynamics of linear quantum open systems}


 
\date{\today}

\begin{abstract}
We study the behavior of networks of quantum oscillators coupled with arbitrary external environments. We analyze the evolution of the quantum state showing that the reduced density matrix of the network always obeys a local master equation with a simple analytical solution. We use this to study the emergence of thermodynamical laws in the long time regime. We demonstrate two main results on thermodynamics:  First, we show that it is impossible to build a quantum absorption refrigerator using linear networks (therefore, such refrigerators require non-linearity as a crucial ingredient, as  proposed by Kosloff and others \cite{Kosloff1}). Then, we show that the third law imposes constraints on the low frequency behavior of the environmental spectral densities. 
\end{abstract}

\maketitle

Deriving the laws of thermodynamics from a quantum substrate is relevant not only for  fundamental but also for practical reasons \cite{KimMahler}. In the macroscopic domain such laws determine the ultimate limits on cooling and work extraction. However, when quantum effects dominate, thermodynamical laws must be derived (not assumed), which is still a controversial issue. In particular, the ultimate limitations on cooling, imposed by the third law have been recently debated \cite{Kosloff1,Kuritzki}. Moreover, a variety of quantum devices have been proposed to act as engines or refrigerators \cite{Kosloff1,others}. The resources required for such machines to operate are not fully known. Among them, the quantum absorption refrigerator \cite{Kosloff1}, whose description does not admit a phenomenological approach.

We study the dynamics and the thermodynamics of arbitrary networks of $N$ oscillators moving in $D$ dimensions while coupled with external reservoirs as shown in Figure 1. This is a generalization of the Quantum Brownian Motion (QBM) model \cite{ZurekPaz,HuPazZhang,FlemingHuRoura,DavilaPaz}. Being the paradigm for an open quantum system, this model has been used to study the emergence of classicality through decoherence \cite{ZurekPaz}. Our work not only has applications to quantum refrigerators but also to other systems where a detailed understanding of heat transport \cite{Dhar} and decoherence is required. This is the case for trapped--ion quantum simulators \cite{Heffnergroup,LinDuan}. It also may help in understanding other natural processes such as the high efficiency of energy transfer in light harvesting complexes  \cite{biomodel}. 

We present four main results. First we show that the dynamics is such that: (i) the quantum state always satisfies a local master equation and (ii) such equation always has a  simple analytical solution. Then we use these results to study the long time limit. We present a simple derivation of thermodynamical laws and prove two new results: (iii) We show that it is impossible to create a quantum absorption refrigerator using linear networks. Such refrigerators have no movable parts and induce the energy to flow away from a cold reservoir by coupling it with a system that is itself coupled with other (hot) reservoirs.  In the quantum regime they were proposed by Kosloff and others \cite{Kosloff1} using a non-linear model. Our work shows that non-linearity is, in fact, an essential ingredient of a quantum absorption refrigerator. Finally, we show that: (iv) the validity of the third law (as stated by Nernst: entropy flow vanishes at zero temperature) imposes a constraint on the environmental spectral density, which for low frequencies must behave as $\omega^\nu$ with $\nu> 0$. 
\begin{figure}[h!]
\centering
\includegraphics[width=1.0\columnwidth]{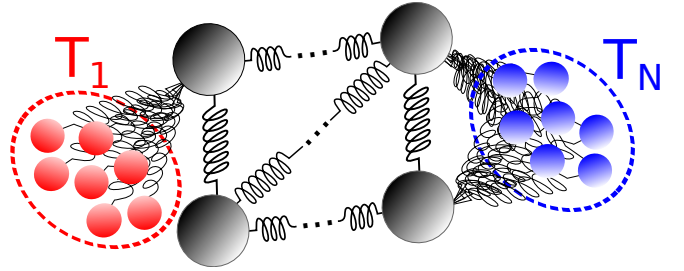}
\caption{We analyze the most general network of interacting oscillators coupled with bosonic reservoirs characterized by arbitrary spectral densities and initial thermal states.}
\label{fig:themodel}
\end{figure}

We consider the total Hamiltonian $H_T=H_S+H_E+H_{int}$, where the system Hamiltonian is $H_S=P^T P/2+X^TVX/2$ ($X$ and $P$ are column vectors storing the $K=ND $ system's coordinates and momenta). The $K\times K$ matrix $V$ defines the network's couplings (we consider unit masses and use superscript $T$ for the transpose). The environmental Hamiltonian is $H_E=\sum_\alpha H_{\alpha}$ with $H_{\alpha}=\sum_k ({\pi^{(\alpha)}_k}^2/2m_k+m_k\omega_k^2 {q^{(\alpha)}_k}^2/2)$. The interaction is $H_{int}=\sum_{\alpha,i,k} C^{(\alpha)}_{ik}x_i q^{(\alpha)}_k$. We study the evolution of the reduced density matrix of the system $\rho= \tr_E(\rho_{SE})$. 

The following two results are valid if the time evolution preserves Gaussian states (i.e., if the total Hamiltonian is quadratic in $P$ and $X$). In this case: (i) $\rho(t)$ satisfies the time-local master equation:
\begin{eqnarray}
        \dot\rho &=& -i[H_R(t),\rho]       
-i\Gamma_{ij}(t)[x_i,\{p_j,\rho\}]\nonumber\\
&-&i\tilde\Gamma_{ij}(t)[p_i,\{x_j,\rho\}] 
-D_{ij}(t)[x_i, [x_j,\rho]]\nonumber\\
&-& \tilde D_{ij}(t)[p_i, [p_j,\rho]] 
-F_{ij}(t)  [x_i,[p_j,\rho]].
\label{mastereq}
\end{eqnarray}
Here $H_R(t)=P^T M_R^{-1}(t) P/2+X^TV_R(t)X/2+f^T(t)X+\tilde f^T(t)P$ is a renormalized Hamiltonian that includes time dependent couplings, masses and forces through $V_R(t)$, $M_R(t)$, $f(t)$ and $\tilde f(t)$. The master equation includes non-unitary effects as relaxation (through $\Gamma(t)$ and $\tilde\Gamma(t)$) and  diffusion (through $D(t)$, $\tilde D(t)$ and $F(t)$). The second result is: (ii) The state $\rho(t)$ can be written as a function of time in a simple way using the characteristic function defined as $\chi(\kappa,t)= \tr (\rho(t) \hat D(\kappa))$ (where the displacement operator is defined as  $\hat D(\kappa)=\exp(-i(P k_p -X k_x))$ and $\kappa=(k_x,k_p)$ is a $2K$--component vector defining momentum and position displacements). This function provides  a complete description of the state and satisfies that 
\begin{equation}
\chi(\kappa,t)=\chi(\Phi(t) \kappa,0) \exp(-{1\over 2}
\kappa^T  \Sigma(t) \kappa)\exp(i \Pi^T(t)\kappa).\label{characteristic}
\end{equation}
The $2K\times 2K$-- matrices $\Phi(t)$ and $\Sigma(t)$ and the $2K$--vector $\Pi(t)$ depend on time. According to (1) and (2), which are valid for arbitrary Gaussian channels, the state evolves through a  combination of a phase space flow and a Gaussian modulation (the environment only induces renormalization, friction and diffusion).  

We sketch here the derivation of equations (1) and (2). To the best of our knowledge they are both new results (details can be found in the supplementary material \cite{SM}). Equation (1)  is a generalization of the master equation obtained for QBM in \cite{HuPazZhang}. Equation (2) is a generalization of a classical result by Rieder, Lebowitz and Lieb (see \cite{Dhar} for references). For QBM eq. (2) was first discussed in \cite{FlemingHuRoura}. Both results follow from properties of the evolution super-operator for $\rho(t)$, which is such that $\rho(t)={\mathcal J}(\rho(0))$. Any super--operator preserving Gaussian quantum states (together with hermiticity and trace) can be written in the position representation as: ${\mathcal J}(z,z',t; z_0,z'_0, 0)\equiv \langle z|{\mathcal J}(|z'_0\rangle\langle z_0|)|z'\rangle = {\rm det}(b_3)\times e^{i(\xi^T b_1 Z+\xi^Tb_2 Z_0+\xi^T_0b_3 Z+\xi^T_0b_4 Z_0)} e^{i(c^T_1 \xi + c^T_2 \xi_0)}\times e^{-\xi^Ta_{1}\xi-\xi^Ta_{2}\xi_0-\xi^T_0a_{3}\xi_0}  /(2\pi)^K$ (here $\xi=z-z'$ and $Z=(z+z')/2$). The matrices $b_m$ and $a_{m}$ together with the vectors $c_m$ parametrize any quantum evolution and depend on the microscopic model, as described  below. Equation (1) is obtained following the method discussed in \cite{ZurekPaz, DavilaPaz}: Computing the time derivative of the propagator we can show that $\dot {\mathcal J}(z,z',t; z_0,z'_0, 0)=P(Z,\xi,Z_0,\xi_0)\times {\mathcal J}(z,z',t; z_0,z'_0, 0)$ where $P$ is a quadratic polynomial of its arguments. All terms in $P\times {\mathcal J}$ that are proportional to the initial coordinates $Z_0$ and $\xi_0$ can be rewritten in a simple way. Thus, using the Gaussian nature of ${\mathcal J}$ we can show that they can be expressed  as a linear combination of terms proportional to $Z$, $\xi$ and to the derivatives of ${\mathcal J}$ with respect to those coordinates. Doing this, the master equation (1) is obtained. Derivation of (2) is even simpler: it follows by integrating over initial coordinates using the Gaussian propagator. 

Matrices in (1) and (2) are determined by those appearing in the propagator ${\mathcal J}$. For the generalized QBM model defined by the Hamiltonian $H_T$ we can compute ${\mathcal J}$ using path integral techniques \cite{FV}. As described in \cite{SM} ${\mathcal J}$ depends only on tw§o properties of the environment: the initial temperatures and the spectral density, defined as $I(\omega)=\sum_\alpha I^{(\alpha)}(\omega)$, where $I^{(\alpha)}_{ij}(\omega)=\sum_k C^{(\alpha)}_{ik}C^{(\alpha)}_{jk} \delta(\omega-\omega_k) /2m_k \omega_k$ (these are real, symmetric and positive $K\times K$ matrices). The initial state of each environment is thermal with temperature $T_\alpha$. These characters appear in the dissipation kernel defined as $\gamma(\tau)=\int_0^\infty d\omega I(\omega) \cos(\omega\tau)/\omega$ and in the noise kernel $\nu(\tau)=\int_0^\infty d\omega \hat\nu(\omega)\cos(\omega\tau)$ (where $\hat\nu(\omega)=\sum_\alpha I^{(\alpha)}(\omega)\coth(\omega/2k_BT_\alpha)$). As shown in \cite{SM}, all coefficients in (1) and (2) depend on the spectral density throught the $K\times K$ matrix $G(t)$, which is the unique solution of:
\begin{equation}
\ddot G + V_R G +2\int_0^t d\tau \gamma(t-\tau) \dot G(\tau)=0,
\label{eqforG}
\end{equation}
with initial conditions $G(0)=0$ and $\dot G(0)=\hat 1$ (the renormalized potential $V_R=V-2\gamma(0)$ is the asymptotic limit of $V_R(t)$ for large values of the cutoff). Equation (3) can be solved using the Laplace transform (the transformed of $G(t)$ is denoted as $\hat G(s)$). Thus, we find $\hat G(s)=(s^2 I + V_R + 2s\hat\gamma(s))^{-1}$, where $\hat\gamma(s)=\int_0^\infty d\omega I(\omega) s^2/(\omega^2+s^2)$ is the transform of the dissipation kernel. All coefficients in (1) are listed in the supplementary material. For example, we find $M(t)=\hat 1$, $f(t)=0=\tilde f(t)$ (i.e., no mass renormalization nor renormalized forces) and $\tilde\Gamma(t)=\tilde D(t)=0$  (i.e. only normal dissipation and no momentum diffusion). Moreover, $V_R(t)=(\ddot G \dot G^{-1} \ddot G - \dot{\ddot G})(\dot G - \ddot G \dot G^{-1} G)^{-1}$ and 
$2\Gamma (t)=(\dot{\ddot G} \dot G^{-1} G - \ddot G)
(\dot G - \ddot G \dot G^{-1} G)^{-1}$. In turn, the solution (2) is such that $\Pi(t)=0$ and the matrices $\Phi(t)$ and $\Sigma(t)$ are:  
\begin{equation}
\Phi(t)=
\begin{bmatrix}
\dot G(t) & G(t)\\
\ddot G(t) & \dot G(t)
\end{bmatrix} {\rm and \ } 
\Sigma(t)= 
\begin{bmatrix}
\sigma^{(0,0)}(t)&\sigma^{(0,1)}(t)\\ \sigma^{(1,0)}(t) & \sigma^{(1,1)}(t)
\end{bmatrix}.
\end{equation}
Here, the $K\times K$ submatrices of $\Sigma(t)$ are $\sigma^{(n,m)}(t)=\int_0^t\int_0^t dt_1 dt_2  G^{(n)}(t_1)\nu(t_1-t_2)G^{(m)}(t_2)$. Diffusion matrices in (1) are also determined by $\Sigma(t)$. For example, $D(t)={\rm Sym}(V_R(t)\sigma^{(0,1)}(t) +2\Gamma(t) \sigma^{(1,1)}(t) +\dot\sigma^{(1,1)}(t))$. 

As seen from (2), the final state is independent of the initial one if and only if $G(t)$ decays (indeed, this is the case because $\chi(0,t)=1$). Assuming that this condition is satisfied we can derive important properties of the stationary regime ($G(t)$ decays for certain spectral densities and coupling matrices but it exhibits revivals for finite reservoirs). To study energy transfer between the  system and the environments we compute  the time derivative of the expectation value of the renormalized Hamiltonian, i.e. $d(\langle H_R\rangle)/dt$. The master equation implies that $d(\langle H_R\rangle)/dt={\rm tr}(D-2\Gamma \sigma^{(1,1)})$. Replacing the formula for the diffusion matrix we get that, in the stationary regime, $d(\langle H_R\rangle)/dt={\rm tr}(V_R \sigma^{(0,1)})$. The trace involves a summation over all sites of the network. We will consider first the case where the network is divided into non--overlapping regions $S_\alpha$ ($\alpha =1,..., R$). We will also assume that each region is coupled with an environment $E_\alpha$ (these assumptions will be relaxed below). In this case we can write $d(\langle H_R\rangle)/dt=\sum_\alpha \dot Q_\alpha$. Here, $\dot Q_\alpha$ is the heat current entering $S_\alpha$ and is given as $\dot Q_\alpha={\rm tr}\left(P_{S_\alpha}V_R \sigma^{(0,1)}\right)$ (where $P_{S_\alpha}$ is the projector onto $S_\alpha$). In fact, $\dot Q_\alpha$ is equal to the mean value of the power transmitted by $E_\alpha$ to the system. In the stationary limit we obtain that the conservation law $\sum_\alpha \dot Q_\alpha=0$ is satisfied. Also, we obtain explicit expressions for the heat currents using that  $\sigma^{(n,m)} \rightarrow \re\int_0^\infty d\omega\omega^{n+m} i^{m-n} \hat G(i\omega) \hat\nu(\omega) \hat G(-i\omega)$.  Then: 
\begin{equation}
\dot Q_\alpha=\sum_\beta \int_0^\infty \omega d\omega \dot {\mathcal Q}_{\alpha\beta}(\omega)\coth(\omega/2k_BT_\beta),
\end{equation}
where 
\begin{eqnarray}
\dot {\mathcal Q}_{\alpha\beta}(\omega)&=& \im {\rm tr}(P_{S_\alpha} V_R \hat G(i\omega) I^{(\beta)}(\omega) \hat G(-i\omega)),\\
&=&-\pi {\rm tr}(I_\alpha(\omega)\hat G(i\omega)I_\beta(\omega)\hat G^\dagger(\omega))\le 0 \quad \alpha\neq \beta\nonumber.
\end{eqnarray}
The inequality in the second line of (6) follows from the positivity of the spectral densities $I_\alpha$ and $I_\beta$. To obtain (6) (see supplementary material) we used the definition of $\hat G(s)$ together with the condition $\re(2\omega\hat\gamma(i\omega))=\pi I(\omega)$ (which is nothing but the simplest form of the fluctuation dissipation theorem \cite{HuPazZhang}). In what follows, we will only need to use the following general properties of the heat transfer matrix $\dot {\mathcal Q}_{\alpha\beta}(\omega)$ that are a direct consequence of (6): i) $\dot {\mathcal Q}_{\alpha\beta}(\omega)=\dot {\mathcal Q}_{\beta\alpha}(\omega)$; ii) $\dot {\mathcal Q}_{\alpha\beta}(\omega)\le 0$ if $\alpha\neq \beta$; iii) $\sum_\alpha \dot {\mathcal Q}_{\alpha\beta}(\omega)=0$. 

Using the above results we can derive the following thermodynamical laws: a) {\em Equilibrium: Energy flows through the system if and only if there is a temperature gradient}. This is due to the fact that the heat flow into the region $S_\alpha$ is  
\begin{equation}
\dot Q_\alpha= -2\sum_{\beta\neq \alpha}\int_0^\infty \omega d\omega \dot {\mathcal Q}_{\alpha\beta}(\omega)(n_\alpha(\omega)- n_\beta(\omega)), 
\end{equation}
where $n_\alpha(\omega)=(\coth(\omega/2k_BT_\alpha)-1)/2$. As $\dot {\mathcal Q}_{\alpha\beta}(\omega)\le 0$, the heat current $\dot Q_\alpha$ vanishes if and only if  $T_\alpha=T_\beta$ for all $\alpha,\beta$. This is equivalent to the 0--th law of thermodynamics: When all the environments have the same temperature then energy does not flow through the system and equilibrium is reached. 

b) {\em Heat flows from the hot to the cold reservoir.} This is Clausius version of the second law of thermodynamics and can be shown as follows:  If $T_\alpha$ is the highest temperature, then $n_\alpha(\omega)>n_\beta(\omega)$ for all $\beta$. Therefore, as $\dot {\mathcal Q}_{\alpha\beta}(\omega)\le 0$ equation (7) implies that $\dot Q_\alpha\ge 0$. Thus, the hottest reservoir always injects energy in the system and this energy is absorbed by the other reservoirs (as implied by the conservation law).  Another formulation of the second law in terms of the entropy flow $\dot S=\sum_\alpha \dot Q_\alpha/ T_\alpha$ states that $S\le 0$, which follows from (7), since $\dot S=-\int_0^\infty d\omega \sum_{\alpha \neq \beta} \dot {\mathcal Q}_{\alpha\beta}(\omega) (1/ T_\beta-1/ T_\alpha) (n_\beta(\omega)-n_\alpha(\omega))\le 0$.

The above argument can be used to show that the coldest reservoir always absorbs energy from the system. In fact, if $T_\alpha<T_\beta$ for all $\beta$ equation (7) implies that  $\dot Q_\alpha\le 0$. For this reason, the class of networks we considered cannot be used to build an absorption refrigerator. Such quantum refrigerators (without movable pieces) were discussed by Kosloff and co-workers \cite{Kosloff1}. Our result implies that absorption refrigerators require non-linearity as an essential ingredient. As stated, the above no-go theorem for linear absorption refrigerators applies for any spectral density. 

The assumptions we made in the above derivation can be relaxed. Thus, we assumed that each region of the system couples with a single environment. We can generalize this as follows: We assume that the region $\alpha$ couples with $N_\alpha$  environments labeled by the index $a_\alpha$ (each having temperature $T_{a_\alpha}$). For simplicity, we also assume that all the spectral densities $I^{(\alpha,a_\alpha)}$ are the same (this can also be relaxed), i.e. $\sum_{a_\alpha} I^{(\alpha,a_\alpha)}(\omega)=N_\alpha I^{(\alpha)}(\omega)$. In this case we can show that the heat flow into the region $S_\alpha$ is given by equation (7), where the temperature dependent factor $(n_\alpha(\omega)-n_\beta(\omega))$ must be replaced by a factor $(\tilde n_\alpha(\omega)-\tilde n_\beta(\omega))$, where $\tilde n_\alpha(\omega)=\sum_{a_\alpha} n_{a_\alpha}(\omega)/N_\alpha$ is the average number of excitations of the environments coupled with $S_\alpha$. In the high temperature limit, the temperature dependent factor in (7) becomes proportional to the difference between the average temperature of the reservoirs, but this factor depends nonlinearly on the temperatures otherwise. Thus, the second law in this case states that the heat always flows into the region that is coupled with the reservoirs with the largest average number of excitations (or the largest average temperature in the high temperature limit). Equivalently, heat always flows away from the region coupled with the reservoirs with the smallest average number of excitations (temperature).  Of course, the environment with the lowest temperature may be coupled with a region which is itself coupled with hotter environments in such a way that the average number of excitations is not the lowest one. Then, heat would flow away from the such environments. However, it is clear that by enlarging the definition of 'the system' and by using the previous argument, one can prove that the environment with the lowest temperature would absorb heat. 

The other assumption in the above derivation is that different regions $S_\alpha$ do not overlap. This can also be generalized using the fact that the obtained results are valid for arbitrary networks. Thus, the case of overlapping regions can be analyzed as follows: Consider a network with regions $S_2\subset S_1$. If $S_{1,2}$ couple with different environments $E_{1,2}$, this situation is physically equivalent to that of a network where $S_1$ couples with $E_1$ while $S_2$ couples strongly with another network $S_2'$ which is itself coupled with $E_2$. In such case, all previous results directly apply. As a consequence of this, the no--go theorem for quantum absorption refrigerators applies to all linear networks.

The above results are valid for arbitrary spectral densities. However, we now show that the validity of the third law imposes constraints on $I(\omega)$.  To see this, we consider environments coupled with single sites, with spectral densities $I^{(j)}_{jj}(\omega)=\gamma_j\omega^{p_j} \theta(\omega)$ ($\theta(\omega)$ is a cutoff function vanishing when $\omega\ge\Lambda$ and $\gamma_j$ a coupling constant). When $p_j=1$ the environment is ohmic while it is super--ohmic (sub--ohmic) for $p_j>1$ ($p_j<1$). Using this, we now prove that: {\em c) The third law emerges if $p_j>0$.}. Thus, the heat flow into the $j$--th site is:
\begin{eqnarray}
\dot Q_j&=& 2\pi\sum_{k\neq j}\int_0^\infty \omega^{1+p_j+p_k}\gamma_j\gamma_k \theta^2(\omega) d\omega |\hat G(i\omega)_{jk}|^2\nonumber \\
&\times& \sum_{a\ge 1}(e^{-a\omega/k_BT_j}-e^{-a\omega/k_BT_k}). \label{explicit}
\end{eqnarray}
For low temperatures the integral is dominated by low frequencies. Taylor expanding the integrand and assuming that all $T_j$ are close to the average $\bar T$ we find:  
\begin{equation}
\dot Q_j= \sum_{k\neq j} \gamma_j\gamma_k {\bar T}^{1+p_j+p_k} (T_j- T_k)(2+p_j+p_k)\alpha_{kj}.
\label{explicit3}
\end{equation}
where $\alpha_{lj}=2\pi (k_B)^{2+p_j+p_k}|\hat G(0)_{jk}|^2\Gamma(2+p_j+p_k)\zeta(2+p_j+p_k)$. Thus, for two reservoirs with the same spectral density (i.e., $p_A=p_B=p$) and a small temperature difference  we find that the heat flow is $\dot Q_A = 2(p+1)\gamma^2 \alpha_{AB} {\bar T}^{2p+1}\Delta T$. This implies that the entropy flow into site--$A$ is  $\dot S_A =\dot Q_A/T_A \propto \bar T^{2p}$. 
Thus, if $p>0$ the entropy flow vanishes when $\bar T\rightarrow 0$, which is nothing but Nernst's (static) version of the third law. 

The results reported here concern the evolution of open networks of quantum oscillators. Thus, in this case: (i) we deduced an exact master equation and (ii) we obtained its explicit solution. We used this to show how thermodynamical laws emerge from first principles (these laws emerge for arbitrary environments whenever $G(t)$ decays for long times). In this case, we proved: (iii) a no-go theorem for linear absorption refrigerators and (iv) we showed that the validity of the third law imposes a constraint on the low frequency behavior of the spectral densities. Very few derivations of the third law are available. Our treatment enabled us to obtain a straightforward derivation of the "static" version of such law (as postulated by Nernst: the entropy flow from any substance is zero at absolute zero). The validity of the dynamical version of the third law (the unattainability of zero temperature in finite time) seems to impose even stronger constraints on the spectral density $I(\omega)$, as recently reported \cite{Kosloff1,Kuritzki}: the static version of the third law is consistent with sub-ohmic,   ohmic and super-ohmic environments whereas the sub-ohmic ones are excluded according to \cite{Kosloff1}. Finally, we point out that it is possible to extend our results to driven systems with quadratic Hamiltonians. This could be done using the above methods, computing time averaged quantities with an approach similar to that of  \cite{Kosloff1,Lili}. 

This work was supported by ANPCyT, CONICET and Ubacyt. E.A.M. is now at Institute fur Experimentalphysik., Uni. Innsbruck,
Technikerstr. 25, A-6020 Innsbruck, Austria.

\section{Supplementary Material}

We present here a detailed description of five important results contained in the main
text. (1): We present the general derivation of the master equation for all systems where 
the Gaussian nature of quantum states is preserved. (ii) We show that for such systems
the characteristic function defined in the text has a simple analytical solution. (iii): We 
discuss in detail the form of the master equation and the exact solution for a microscopic 
model: the generalized quantum Brownian motion. (iv): We present a derivation of the 
equations (5), (6) and (7) of the main text providing an explicit formula for the heat transfer 
matrix. (v): We prove the generalized fluctuation dissipation relation used in the above 
derivations. 

\subsection{1) Derivation of equation (1) of the main text: The general master equation.}
 
The master equation (1) and its exact solution (2) are a consequence of a single assumption: the evolution preserves the Gaussian nature of states. This is equivalent to assuming that the total Hamiltonian is a quadratic form of position and momenta (both of system and environments) and that the initial state of the environments are Gaussian (valid for thermal states). To preserve Gaussian states the evolution operator for the reduced density matrix of the system must satisfy certain simple properties. Thus, such operator is such that $\rho(t)={\mathcal J}(\rho(0))$. To preserve Gaussian states the matrix elements ${\mathcal J}(x,x',t;x_0,x'_0,0)=\langle z|{\mathcal J}(|x_0\rangle\langle x'_0|)|x' \rangle$ must be a Gaussian function of its arguments. Using sum and difference coordinates $Z=(x+x')/2$ and $\xi=x-x'$, the most general quadratic form is:
\begin{align}
\label{eq:supp-j}
&{\mathcal J}(Z,\xi,t; Z_0, \xi_0, 0) \equiv {\mathcal N} e^{i(\xi^T b_1 Z+\xi^Tb_2 Z_0+\xi^T_0b_3 Z+\xi^T_0b_4 Z_0)} \times \nonumber\\
&\times e^{i(c^T_1 \xi + c^T_2 \xi_0)} e^{-\xi^Ta_{1}\xi-\xi^Ta_{2}\xi_0-\xi^T_0a_{3}\xi_0}\times\\
&\times e^{i(d^T_1 Z + d^T_2 Z_0)} e^{-Z^Te_{1}Z-Z^Te_{2}Z_0-Z^T_0e_{3}Z_0}.\nonumber
\end{align} 
Here, the time-dependent matrices $b_m$, $a_{m}$ and $e_{m}$ together with the time dependent vectors $c_l$ and $d_m$ and the normalization function ${\mathcal N}$ parametrize any quantum evolution preserving Gaussian states. These matrices are constrained by two simple conditions. First, the hermitian nature of quantum states must be conserved. This requires that 
${\mathcal J}(Z,-\xi,t; Z_0, -\xi_0,0)^*={\mathcal J}(Z,\xi,t; Z_0, \xi_0,0)$. This implies that all the above matrices must be real. In turn, the preservation of the trace implies that $\int dZ {\mathcal J}(Z,\xi=0,t,Z_0,\xi_0,0)=\delta(\xi_0)$. As a consequence of this, it is simple to show that $e_{n}$ and $d_n$ must vanish and that the normalization must be given as ${\mathcal N}={\rm det}(b_3)/(2\pi)^K$. Therefore, any propagator preserving Gaussianity, hermiticity and trace reads:
\begin{align}
\label{eq:jvsabc}
&{\mathcal J}(Z,\xi,t; Z_0, \xi_0, 0) = \frac{\det(b_3)}{(2\pi)^K}\times\nonumber\\
&\times e^{i(\xi^T b_1 Z+\xi^Tb_2 Z_0+\xi^T_0b_3 Z+\xi^T_0b_4 Z_0)} \times \nonumber\\
&\times e^{i(c^T_1 \xi + c^T_2 \xi_0)} e^{-\xi^Ta_{1}\xi-\xi^Ta_{2}\xi_0-\xi^T_0a_{3}\xi_0}.
\end{align} 
The time dependent matrices $a_n$, $b_n$ and $c_n$ are determined by the microscopic model. Below, we will describe this for the specific model analyzed in the main text: the generalized quantum Brownian motion model. However, to obtain the master equation (1) and the exact solution (2) it is not necessary to give the specific form of these matrices. Thus, the master equation is obtained in the following way: We can use
first compute the time derivative of the evolution operator ${\mathcal J}$ using equation \ref{eq:jvsabc}. 
By doing this we find that $\dot {\mathcal J}$ can be written as a second degree polynomial multiplied by ${\mathcal J}$ itself. Thus, 
\begin{equation}
\label{eq:i-JPJ}
\frac{\partial {\mathcal J}}{\partial t}
= P(Z,\xi,Z_0,\xi_0){\mathcal J}, 
\end{equation}
where the polynomial $P(Z,\xi,Z_0,\xi_0)$ is
\begin{align}
\label{eq:supp-time-d}
&P(Z,\xi,Z_0,\xi_0)
 =  \frac{\det(\dot{b}_3)}{\det(b_3)} + i(\xi^T \dot{b}_1 Z+\xi^T \dot{b}_2 Z_0+\xi^T_0 \dot{b}_3 Z\nonumber\\
& +\xi^T_0 \dot{b}_4 Z_0) + i(\dot{c}^T_1 \xi + \dot{c}^T_2 \xi_0) -\xi^T \dot{a}_{1}\xi-\xi^T \dot{a}_{2}\xi_0-\xi^T_0 \dot{a}_{3}\xi_0 .
\end{align}

Then, we use again equation \ref{eq:jvsabc} to show that all the terms proportional to $Z_0 {\mathcal J}$ and $\xi_0 {\mathcal J}$ can be rewritten as a linear combination of others which are proportional to the final coordinates $Z$ and $\xi$ and the derivatives of the propagator with respect to them. For instance, the linear terms can be written as:
\begin{align}
\label{eq:supp-xi-0}
\xi_0^T {\mathcal J} &= -i \frac{\partial {\mathcal J}}{\partial Z} b_3^{-1} - \xi^T b_1 b_3^{-1} {\mathcal J},\\
\label{eq:supp-z-0}
Z_0^T {\mathcal J} &= -i \frac{\partial {\mathcal J}}{\partial \xi} b_2^{-T} - z^T b_1^T b_2^{-T} {\mathcal J} - c_1^T b_2^{-T} {\mathcal J} \nonumber \\
& \quad - \frac{\partial {\mathcal J}}{\partial Z} b_3^{-1} a_2^T b_2^{-T} + i \xi^T (b_1 b_3^{-1} a_2^T - 2 a_1) b_2^{-T} {\mathcal J},
\end{align}
and likewise with the quadratic terms (not shown here). By grouping terms in (\ref{eq:supp-time-d}) according to their dependence on ${\mathcal J}$ and its derivatives we can obtain the master equation in the $Z, \xi$ coordinates:
\begin{align}
\frac{\partial \rho}{\partial t} &= -i  \nabla_{\xi}^T M_R \nabla_Z^{}  \rho(Z, \xi, t) - i \xi^T V_{R}^{}(t) Z \, \rho(Z, \xi, t) \nonumber \\
& \quad - i \xi^T f(t) \rho(Z, \xi, t) -  
\nabla_\xi^T \tilde f(t) \rho(Z, \xi, t)\nonumber\\ 
& \quad - 2 \xi^T \Gamma(t) \nabla_\xi^{} \rho(Z, \xi, t) -
 2 \nabla_Z^T \tilde\Gamma(t) {Z} \rho(Z, \xi, t)\nonumber\\ 
&- \xi^T D(t) \xi \, \rho(Z, \xi, t)
- i \nabla_Z^{T} F(t) \xi \, \rho(Z, \xi, t) \nonumber\\
& \quad - \nabla_Z^T \tilde D(t) \nabla_Z^{} \, \rho(Z, \xi, t).
\end{align}
The matrices appearing in this equation depend on $a_m$, $b_m$ and $c_m$ in a rather cumbersome way. These expressions can be read from equations (\ref{eq:supp-xi-0})-(\ref{eq:supp-z-0}). We do not include these formulae here. Instead, we will give the exact expressions for these coefficients for the generalized quantum Brownian motion model. Finally, we can obtain the master equation in operator form using that:
\begin{align}
\langle z|[x, \rho]|z'\rangle &= - \xi \rho(Z, \xi),\quad
\langle z|\{x, \rho\}|z'\rangle = 2 Z \rho(Z, \xi),\nonumber\\
\langle z|[p, \rho]|z'\rangle &= - i \nabla_Z^T \rho(Z, \xi),\quad 
\langle z|\{p, \rho\}|z'\rangle = 2 i \nabla_{\xi}^T \rho(Z, \xi).\nonumber
\end{align}
In this way we find the master equation descirebed in the main text, which reads:
\begin{eqnarray}
        \dot\rho &=& -i[H_R(t),\rho]       
-i\Gamma_{ij}(t)[x_i,\{p_j,\rho\}]\nonumber\\
&-&i\tilde\Gamma_{ij}(t)[p_i,\{x_j,\rho\}] 
-D_{ij}(t)[x_i, [x_j,\rho]]\nonumber\\
&-& \tilde D_{ij}(t)[p_i, [p_j,\rho]] 
-F_{ij}(t)  [x_i,[p_j,\rho]].
\label{mastereq}
\end{eqnarray}
Here $H_R(t)=P^T M_R^{-1}(t) P/2+X^TV_R(t)X/2+f^T(t)X+\tilde f^T(t)P$ is a renormalized Hamiltonian that includes time dependent couplings, masses and forces through $V_R(t)$, $M_R(t)$, $f(t)$ and $\tilde f(t)$. Using this equation it is rather straightforward to derive evolution equation for the first and second moments of position and momentum. Doing this, we can give a simple interpretation for each of the terms appearing in the master equation. Thus, we realize that the evolution of the first moments is fully determined by $H_R$,  $\Gamma(t)$ and $\tilde \Gamma(t)$ (which are time dependent relaxation rates). The second moments also depend on the diffusive matrices $D(t)$, $\tilde D(t)$ and $F(t)$. 

\subsection{2) Derivation of equation (2) of the main text: the quantum state as a function of time}

The characteristic function defined in the main text is $\chi(\kappa)={\rm Tr}(\rho D(\kappa)$, where the $2$--K component vector $\kappa$ is $\kappa=(k,\xi)$ and the  displacement operator is $D(\kappa)=\exp(i \xi^T P + i k^T X)$. It can be shown that this function is nothing but the Fourier transform of the density matrix with respect to the sum coordinate $Z$. Thus, 
\begin{align}
\label{eq:chidef}
\chi(k, \xi, t) = \int dZ \, e^{i k^T Z} \rho(Z, \xi, t).
\end{align}

To obtain $\chi(\kappa)$ at arbitrary times we use that the density matrix at arbitrary times is 
\begin{align}
\label{eq:supp-evo-rho}
\rho(Z, \xi, t) = \int dZ_0 \, d\xi_0 \, {\mathcal J}(Z, \chi, t, Z_0, \chi_0, 0) \rho(Z_0, \chi_0, 0).
\end{align}
We will obtain the explicit form of the evolution operator of the characteristic function. This operator, denoted as ${\mathcal J}_\chi$, is such that 
\begin{align}
\label{eq:supp-evo-chi}
\chi(\kappa, t) & = \int d\kappa_0 \, {\mathcal J}_\chi(\kappa, t, \kappa_0, 0) \chi(\kappa_0, 0),
\end{align}
We can obtain ${\mathcal J}_\chi$ by replacing \ref{eq:supp-evo-rho} in \ref{eq:chidef}, using the explicit form of the propagator \ref{supp-evo-rho} and expressing the initial density matrix in terms of the initial characteristic function. Thus, if we do this we first obtain that 
\begin{align}
\label{eq:supp-j-chi}
{\mathcal J}_\chi(k, \xi, t, k_0, \xi_0, 0) &=  \int dZ \, dZ_0 \, e^{i k^T Z} e^{-i k_0^T Z_0} \times \nonumber \\
& \qquad \times J(Z, \xi, t, Z_0, \xi_0, 0).
\end{align}
Then we can use \ref{eq:supp-j} and perform both integrals in \ref{supp-j-chi}. Doing this, it is immediate to see that:
\begin{align}
\label{eq:supp-j-chi-final}
{\mathcal J}_\chi(\kappa,t; \kappa_0,0) &= e^{i \pi^T \kappa} e^{- \kappa^T \Sigma \kappa / 2} \, \delta(\kappa_0 - \Phi^T \kappa), 
\end{align}
where the matrices $\Phi$, $\Sigma$ and $\pi$ depend on time and will be given below in terms of $a_m$, $b_m$ and $c_m$. Using this expression for the propagator we find that the characteristic function is  
\begin{align}
\label{eq:equation-2}
\chi(\kappa, t) &= e^{i \pi^T \kappa} e^{- \kappa^T \Sigma \kappa / 2} \chi(\Phi^T \kappa,0).
\end{align}
This is the equation (2) appearing in the main text. The  time-dependent matrices and vectors appearing in \ref{eq:equation-2} are:
\begin{align}
\Phi = \begin{pmatrix}
-b_3^{-1} b_4 & \quad -b_3^{-1}\\
b_2 - b_1 b_3^{-1} b_4 & \quad -b_1 b_3^{-1}
\end{pmatrix},
\end{align}
\begin{align}
\pi = \begin{pmatrix}
-c_2 b_3^{-T}\\
c_1 - c_2 b_3^{-T} b_1^T\\
\end{pmatrix}, \quad
\Sigma = \begin{pmatrix}
\sigma^{(0,0)} & \sigma^{(0,1)}\\
\sigma^{(1,0)} & \sigma^{(1,1)}
\end{pmatrix},
\end{align}
with:
\begin{align}
\sigma^{(0,0)} &= 2 b_3^{-1} a_3 b_3^{-T},\\
\sigma^{(0,1)} &= \sigma^{(1,0)T} = a_1 + b_3^{-1} (a_2^T + 2 a_3 b_3^{-T} b_1),\\
\sigma^{(1,1)} &= 2 a_1 + 2 \operatorname{Sym} (b_1 b_3^{-1} a_2^T) + 2 b_1 b_3^{-1} a_3 b_3^{-T} b_1.
\end{align}
Below, we will give the explicit form of these matrices for the generalized QBM model. 

\subsection{3) Generalized Quantum Brownian Motion (QBM) model: Exact computation of the time dependent coefficients of equations (1) and (2) of the main text}

As discussed in the main text, we consider the generalized QBM model where the total Hamiltonian is $H_T=H_S+H_E+H_{int}$. The system's Hamiltonian is $H_S=P^T P/2+X^TVX/2$ ($X$ and $P$ are column vectors storing the $K=ND$ system's coordinates and momenta). The $K\times K$ matrix $V$ defines the network's couplings (we consider unit masses). The environmental Hamiltonian is $H_E=\sum_\alpha H_{\alpha}$ with $H_{\alpha}=\sum_k ({\pi^{(\alpha)}_k}^2/2m_k+m_k\omega_k^2 {q^{(\alpha)}_k}^2/2)$. The interaction is $H_{int}=\sum_{\alpha,i,k} C^{(\alpha)}_{ik}x_i q^{(\alpha)}_k$. We study the evolution of the reduced density matrix of the system $\rho$ which is obtained from the full density matrix by tracing out the environment, i.e., $\rho= \tr_E(\rho_{T})$. As we will use path integral techniques in our derivation, we will work with the action that in the model can be split in the same way as the Hamiltonian. Thus, 
$S_T[X,Q]=S_S[X]+S_E[Q]+S_{int}[X,Q]$, where $Q$ is a vector collecting all the coordinates of all the environmental oscillators. 

The evolution operator of the total density matrix is such that $\rho_T(t)={\mathcal K}(\rho_T(0))$. We can write the position representation of such operator as a double path integral over the system and the environment. In fact, if we denote ${\mathcal K}(X_f,Q_f,X'_f,Q'_f, t;X_i,Q_i,X'_i,Q'_i,0)=\langle X_f,Q_f|{\mathcal K}(|X_i,Q_i\rangle\langle X'_i,Q'_i|)|X'_f,Q'_f\rangle$ we can write
\begin{align}
&{\mathcal K}(X_f,Q_f,X'_f,Q'_f,t;X_i,Q_i,X'_i,Q'_i,0)=\nonumber\\ 
&\int DX DQ e^{iS_T[X,Q]} 
\int DX' DQ' e^{-iS_T[X',Q']},\nonumber
\end{align}
where the functional integral involves all trajectories satisfying the boundary conditions $X(0)=X_i$, $Q(0)=Q_i$, $X(t)=X_f$ and $Q(t)=Q_f$ (and the same for the primed variables). 

We assume that the initial state is such that $\rho_T(0)=\rho(0)\otimes \rho_E(0)$. In this case, the evolution operator of the reduced density matrix also has a simple path integral representation. We denote the matrix elements of such operator in the position basis as ${\mathcal J}(X_f,X'_f,t;X_i,X'_i,0)=\langle X_f|{\mathcal J}(|X_i\rangle\langle X'_i|)|X'_f\rangle$. These matrix elements have the following path integral representation:
\begin{align}
\label{eq:reducedpropagator}
&{\mathcal J}(X_f,X'_f,t;X_i,X'_i,0)= 
\int DX  e^{iS_S[X]} \nonumber\\
&\times \int DX'
 e^{-iS_S[X']}\quad {\mathcal F}[X,X']. 
\end{align}
Here, ${\mathcal F}[X,X']$ is the so--called Feynman--Vernon influence functional whose explicit form is obtained by collecting all the integrals involving environmental quantities. Thus, the influence functional is:
\begin{align}
\label{eq:Influence}
&{\mathcal F}[X,X']=\int dQ_f\int dQ_i\int dQ'_i \rho_E(Q_i,Q'_i,0)\int_{Q_0}^{Q_f} DQ\nonumber\\
&\times \int_{Q'_0}^{Q_f} DQ'
e^{iS_E[Q]+iS_{int}[X,Q]} e^{-iS_E[Q']-iS[X',Q']}. 
\end{align}
For the problem we are considering the influence functional ${\mathcal F}[X,X']$ can be computed because all the integrals appearing in \ref{eq:Influence} are Gaussian. Not surprisingly, the result is also a Gaussian functional of $X$ and $X'$, which is simpler to express in terms of the sum and difference coordinates: $Z=(X+X')/2$ and $\xi=X-X'$. Thus, 
\begin{align}
\label{eq:InfluenceQBM}
&{\mathcal F}[Z,\xi]=\exp(2i\int_0^t dt_1\int_0^{t_1} dt_2 \xi^T(t_1) \eta(t_1-t_2) Z(t_2)\nonumber\\
&-\int_0^t dt_1 \int_0^t dt_2 \xi^T(t_1) \nu(t_1-t_2) \xi(t_2)),
\end{align}
where the dissipation and the noise kernels are:
\begin{align}
\label{eq:kernels}
&\eta(t)=\int_0^\infty d\omega I(\omega)
\sin(\omega t)\nonumber\\
&\nu(t)=\int_0^\infty d\omega \cos(\omega t)
\sum_\alpha I^{(\alpha)}(\omega) \coth(\omega/2k_BT_\alpha),
\end{align}
Here, the spectral densities are $I(\omega)=\sum_\alpha I^{(\alpha)}(\omega)$ where  
\begin{align}
\label{eq:spectraldensity}
I^{(\alpha)}_{ij}(\omega) = \sum_{k = 1}^{M_\alpha} \delta(\omega - \omega_{k}) \frac{C^{(\alpha)}_{i k} C^{(\alpha)}_{j k}}{2 m_{k} \omega_{k}}.
\end{align}

Using these results, we can write the path integral representation of the evolution operator for the reduced density matrix using sum and difference coordinates as:
\begin{equation}
\label{eq:propagatorsumdif}
{\mathcal J}(Z_f,\xi_f,t;Z_i,\xi_i,0)= 
\int DZ  \int D\xi \quad  e^{iA[Z,\xi]}, 
\end{equation}
where the effective action is
\begin{align}
\label{eq:efactionfirst}
&A[Z,\xi]=\int_0^t dt_1(\dot\xi^T(t_1)\dot Z(t_1) -  \xi^T(t_1) V Z(t_1))\nonumber\\
&+2\int_0^t dt_1\int_0^{t_1} dt_2 \xi^T(t_1) \eta(t_1-t_2)Z(t_2)\nonumber\\
&+i\int_0^t dt_1 \int_0^t dt_2 \xi^T(t_1) \nu(t_1-t_2) \xi(t_2)).
\end{align}
It is convenient to integrate this by parts to obtain:
\begin{align}
\label{eq:efactionsecond}
&A[Z,\xi]=(\xi^T(t_1) \dot Z (t_1)|_0^t+
\int_0^t dt_1\xi^T(t_1)(\ddot Z(t_1) - V_R Z(t_1))\nonumber\\
&+2\int_0^t dt_1\int_0^{t_1} dt_2 \xi^T(t_1) \gamma(t_1-t_2)\dot Z(t_2)\nonumber\\
&+i\int_0^t dt_1 \int_0^t dt_2 \xi^T(t_1) \nu(t_1-t_2) \xi(t_2)),
\end{align}
where the renormalized potential is $V_R=V-2\gamma(0)$ and the kernel $\gamma(t)$ is defined as
\begin{equation}
\label{eq:gamma}
\gamma(t)=\int_0^\infty d\omega {\frac{I(\omega)}{\omega}} \cos(\omega t).
\end{equation} 
The path integral in \ref{eq:propagatorsumdif} can be computed by integrating over  $\delta Z$ and $\delta\xi$ where $Z=Z_c+\delta Z$ and $\xi=\xi_c+\delta\xi$ being $Z_c$ and $\xi_c$ the trajectories that extremize the real part of $A[Z,\xi]$. The equation defining $Z_c$ is 
\begin{equation}
\label{eq:xic}
\ddot Z_c(t)+V_R Z_c(t)+2\int_0^t dt_1\gamma(t-t_1)\dot Z_c(t_1)=0. 
\end{equation}
In turn, $\xi_c(\tau)$ is such that $\xi_c(\tau)=Z_c(t-\tau)$. Then,
\begin{align}
\label{eq:propagatoclassical}
{\mathcal J}(Z_f,\xi_f,t;Z_i,\xi_i,0)=e^{iA[Z_c,\xi_c]} \int D\delta Z  D\delta \xi \quad e^{iA[\delta Z,\delta \xi]}, 
\end{align}
All the dependence on initial and final coordinates is contained in $A[Z_c,\xi_c]$ since the remaining path integral involves functions that vanish at the initial and final times.  We remark that we obtained equation \ref{eq:propagatoclassical} even though $Z_c$ and $\xi_c$ extremize only the real part of $A[Z,\xi]$. When doing this, a linear term coupling $\xi_c$ and $\delta\xi$ arising from the imaginary part of $A[Z,\xi]$. However, it can be shown that this term does not contribute to the final result (because the action does not contain terms that are quadratic in $Z$). Using \ref{eq:propagatoclassical}, the effective action \ref{eq:efactionsecond} and the equation \ref{eq:xic}, we get find that the evolution operator is:
\begin{align}
\label{eq:propagatoalmostfinal}
&{\mathcal J}(Z_f,\xi_f,t;Z_i,\xi_i,0)={\mathcal N}(t) 
e^{i \xi^T_(\tau) \dot Z_c(\tau)|_0^t} \nonumber\\
&\times e^{-\int_0^t dt_1\int_0^t dt_2 \xi^T_c(t_1)\nu(t_1-t_2) \xi_c(t_2)},  
\end{align}
where ${\mathcal N}(t)$ is fixed by normalization. 

To obtain the final result, we need $Z_c(\tau)$. Thus, we must solve the linear, second order integro--differential equation \ref{xic} with the corresponding bounday conditions. The space of solutions has a basis formed by two independent functions that can be chosen as $G_i(\tau)$ and $G_f(\tau)$ satisfying the boundary conditions $G_i(0)=G_f(t)=1$ and $G_i(t)=G_f(0)=0$, we can write:
\begin{equation}
\label{eq:xic2}
Z_c(\tau)= G_i(\tau) Z_i +  G_f(\tau) Z_f. 
\end{equation}
Replacing this in \ref{eq:propagatoalmostfinal}, we obtain the evolution operator as:
\begin{align}
\label{eq:propagatofinal}
&{\mathcal J}(Z_f,\xi_f,t;Z_i,\xi_i,0)={\mathcal N}(t) 
e^{-\xi^T_f a_1 \xi_f -\xi^T_f a_2\xi_i - \xi^T_i a_3 \xi_i}\nonumber\\
&\times
e^{i \xi^T_f b_1 Z_f +i \xi^T_f b_2 Z_i +
        i \xi^T_i b_3 Z_f  +i \xi^T_i b_4 Z_i},
\end{align}
where 
\begin{align}
b_1(t) &= \dot G_f(t),\quad b_2(t) = \dot G_i(t),\\
b_3(t) &= - \dot G_f(0),\quad
b_4(t) = - \dot G_i(t),
\end{align}
and
\begin{align}
a_1(t) &= \frac{1}{2} \int_{0}^{t} dt_1 \int_{0}^{t} dt_2 \, G_i(t_1)^T \nu(t_1 - t_2) G_i(t_2),\\
a_2(t) &= \int_{0}^{t} dt_1 \int_{0}^{t} dt_2 \, G_i(t_1)^T \nu(t_1 - t_2) G_f(t_2), \\
a_3(t) &= \frac{1}{2} \int_{0}^{t} dt_1 \int_{0}^{t} dt_2 \, G_f(t_1)^T \nu(t_1 - t_2) G_f(t_2). 
\end{align}
Finally, it is convenient to express $G_i(t)$ and $G_f(t)$ in terms of $G(t)$ which is a solution of \ref{eq:xic} satisfying the boundary conditions $G(0)=0$ and $\dot G(0)=1$. Thus, 
\begin{align}
G_i(\tau) &= \dot{G}(\tau) - G(\tau) G^{-1}(t) \dot{G}(t),\\
\label{eq:g-f}
G_f(\tau) &= G(\tau) G^{-1}(t).
\end{align}
For example, the coefficients $b_m$ are
\begin{align}
\label{eq:bvsg}
b_1 &= \dot G(t) G^{-1}(t).\quad
b_2 = \ddot{G}(t) - \dot G(t) G^{-1}(t) \dot{G}(t),\\
b_3 &= - G^{-1}(t).\qquad
b_4  =  G^{-1}(t) \dot{G}(t),\
\end{align}
Using this, we obtain that the matrices appearing in the exact solution for $\chi(\kappa,t)$ (equation (2) of the main text) are
\begin{align}
\Phi = \begin{pmatrix}
\dot G(t) & G(t) \\
\ddot{G}(t) & \dot G
\end{pmatrix},\quad
\Sigma = \begin{pmatrix}
\sigma^{(0,0)} & \sigma^{(0,1)}\\
\sigma^{(1,0)} & \sigma^{(1,1)}
\end{pmatrix}.
\end{align}
with
\begin{equation}
\sigma^{(n,m)}(t) =\int_0^t dt_1 \int_0^t dt_2 G^{(n)}(t_1)\nu(t_1-t_2) G^{(m)}(t_2).
\end{equation}
We can also express the matrices appearing in the master equation \ref{eq:mastereq} in terms of the matrix $G(t)$ and $\sigma^{(n,m)}$. They are obtained after some algebra and read as follows:
\begin{align}
V_R(t)&=(\ddot G \dot G^{-1} \ddot G - \dot{\ddot G})(\dot G - \ddot G \dot G^{-1} G)^{-1},\nonumber\\ 
2\Gamma (t)&=(\dot{\ddot G} \dot G^{-1} G - \ddot G)
(\dot G - \ddot G \dot G^{-1} G)^{-1}\nonumber\\
D(t)&={\rm Sym}(V_R(t)\sigma^{(0,1)}(t) +2\Gamma(t) \sigma^{(1,1)}(t) +\dot\sigma^{(1,1)}(t)),\nonumber\\
F(t)&=\sigma^{(1,1)}(t)-\sigma^{(0,0)}(t)V_R(t)-2\sigma^{(0,1)}(t) \Gamma^T(t)\nonumber\\
&-\dot\sigma^{(0,1)}(t).\nonumber
\end{align}

\subsection{4) Derivation of equations (5), (6) and (7) of the main text: Properties of the heat transfer matrix}

As mentioned in the text, if a stationary state is attained in the long time limit, the time derivative of the expectation value of renormalized Hamiltonian can be written as $d\langle H_R\rangle/dt ={\rm tr}(V_R \sigma^{(0,1)})$, where $V_R$ is the asymptotic value of the renormalized potential and the trace involves a $K\times K$ matrix. This equation can be derived in various ways. The most natural one seems to be the one mentioned in the main text: We can notice that $\langle H_R\rangle =Tr(\rho H_R)$ and use the master equation to show that in the most general case $d\langle H_R\rangle/dt ={\rm tr}(D-2\Gamma \sigma^{(1,1)})$. Now we can use the explicit expression for the relaxation and diffusion matrices ($\Gamma$ and $D$, which are given above) to obtain the final expression $d\langle H_R\rangle/dt =tr(V_R \sigma^{(0,1)})$ which is valid in the asymptotic limit (when all elements of $\Sigma$ become time independent). As it is done in the main text, we first discuss the simplest case and assume that the network is divided into non--overlapping regions $S_\alpha$ each one of which is coupled with a different environment $E_\alpha$. In this case the time derivative of the expectation value of the Hamiltonian can be written as $d\langle H_R\rangle/dt =\sum_\alpha \dot Q_\alpha$, where $\dot Q_\alpha={\rm tr}(P_{S_\alpha} V_R \sigma^{(0,1)})$ being $P_{S_\alpha}$ the projection operator onto the region $S_\alpha$ (these projection operators are $K\times K$ matrices satisfying $\sum_{S_\alpha} P_{S_\alpha}=\hat 1$). The quantity $\dot Q_\alpha$ is the energy that enters the network through the region $S_\alpha$. Moreover, in the asymptotic limit the total energy is conserved and we obtain the conservation law $\sum_\alpha \dot Q_\alpha=0$.  

It is simple to see that using all the above expressions we can derive simple formulae for the heat current $\dot Q_\alpha$. For this, we use that in the long time limit the correlation matrix $\sigma^{(0,1)}$ can be  written in terms of the Laplace transform of $G(t)$ (denoted as $\hat G(s)$), which satisfies:
\begin{equation}
\label{laplaceG}
(s^2 \hat I+V_R+2s\gamma(s))\hat G(s)=\hat I. 
\end{equation}
In fact, in the asymptotic limit we can write 
$\sigma^{(0,1)}_{F} \rightarrow \im\int_0^\infty d\omega \, \omega \hat G(i\omega) \hat\nu(\omega) \hat G^\dagger(i\omega)$, where $\hat\nu(\omega)=\sum_\beta I^{(\beta)}(\omega) \coth(\omega/2k_BT_\beta)$. Using this equation we find that the heat current $\dot Q_\alpha$ is 
\begin{align}
\dot Q_\alpha &= \sum_\beta \im \int_0^\infty \omega \, d\omega \, \coth(\omega/2 k_B T_\beta) \nonumber\\
& \qquad \times {\rm tr}(P_{S_\alpha} V_R \hat G(i\omega) I^{(\beta)}(\omega)\hat G^\dagger(i\omega)).
\end{align}
From this expression it is evident that the heat current $\dot Q_\alpha$ is a sum over all the environments $E_\beta$. Thus, we have  
\begin{equation}
\label{eq:eq5}
\dot Q_\alpha=\sum_{\beta} 
\int_0^\infty \omega d\omega \dot {\mathcal Q}_{\alpha\beta}(\omega)\coth(\omega/2k_BT_\beta).
\end{equation}
where the heat transfer matrix $\mathcal{Q}$ is defined as
\begin{equation}
\dot {\mathcal Q}_{\alpha\beta}(\omega)= \im {\rm tr}(P_{S_\alpha} V_R \hat G(i\omega) I^{(\beta)}(\omega) \hat G^\dagger(i\omega)). 
\end{equation}
In this way we have obtained equation (5) and the first line of equation (6). 
 
To derive the second line of equation (6), obtaining in this way a more convenient expression for the heat transfer matrix, we need the following three steps: The first step is to use the definition of $\hat G(i\omega)$ as \ref{eq:laplaceG} which implies that 
$V_R \hat G(i\omega)= \hat 1 - \omega^2 \hat G(i\omega)-2i\omega \gamma(i\omega)\hat G(i\omega)$. Replacing this in the expression for the heat transfer matrix we find that $\dot {\mathcal Q}_{\alpha \beta}$ is written as the sum of three terms. However, it is simple to show that the first two of them vanish. In fact, the contribution coming from the identity matrix $\hat 1$ vanish since, as regions $S_\alpha$ do not overlap,  $P_{S_\alpha} I^{(\beta)}=0$ when $\alpha\neq \beta$. In turn, the contribution of the term $\omega^2\hat G(i\omega)$ also vanishes because it  turns out to be equal to the trace of the product of the symmetric matrix $P_{S_\alpha}$ and the anti--symmetric matrix $\im \hat G(i\omega) I^{(\beta)}(\omega) \hat G^\dagger(i\omega)$. Therefore, the heat transfer matrix, after the first step of the derivation, turns out to be:
\begin{equation}
\dot {\mathcal Q}_{\alpha\beta}(\omega)= -2\omega \re {\rm tr}(P_{S_\alpha} \gamma(i\omega) \hat G(i\omega) I^{(\beta)}(\omega) \hat G^\dagger(i\omega)). 
\end{equation}
The second step is to notice that, as the matrix $\hat G I^{(\beta)} \hat G^\dagger$ is hermitian, the result is proportional to the real part of the kernel $\gamma(i\omega)$, which is related to the spectral density by the equation $\re(2\omega\gamma(i\omega))=\pi I(\omega)$.  We will prove this relation below. We remark here that this is nothing but the simples version of the fluctuation dissipation relation. Using this, the heat transfer matrix turns out to be:
\begin{equation}
\dot {\mathcal Q}_{\alpha\beta}(\omega)= -\pi {\rm tr}(I^{(\alpha)}(\omega) \hat G(i\omega) I^{(\beta)}(\omega) \hat G^\dagger(i\omega)), 
\end{equation}
if $\alpha \neq \beta$. This is the second line of equation (6) in the main text. From this equation the symmetric nature of the heat transfer matrix is evident. 

Finally, we must show the negativity of the off-diagonal elements of the heat transfer matrix. For this we will use the above expression and take a third and last step. We will use the fact that all spectral densities $I^{(\beta)}(\omega)$ are real, symmetric and positive $K\times K$ matrices. These conditions directly follow from the definition of such spectral densities shown in 
\ref{eq:spectraldensity}. As $I^{(\alpha)}$ is positive (a necessary condition for the stability of the system), all its eigenvalues $\lambda$ are positive. Writing the trace appearing in the expression for $\dot {\mathcal Q}_{\alpha,\beta}$ in the basis of eigenvectors of $I^{(\alpha)}$ (denoted as $\ket{v_{\lambda}}$), we see that 
\begin{align}
\dot {\mathcal Q}_{\alpha\beta}(\omega) &
= -\pi \sum_{\lambda} \lambda 
\langle v_{\lambda}|\hat G(i\omega) I^{(\beta)}(\omega) \hat G^\dagger(i\omega)|v_{\lambda}\rangle .
\end{align}
The above sum is clearly positive due to the positivity of all eigenvalues $\lambda$ and also due to the positivity of $I^{(\beta)}$. In this way, we have proven that $\dot {\mathcal Q}_{\alpha\beta}(\omega) \leq 0$ for $\alpha \neq \beta$.

In turn, the equation (7) of the main text is obtained 
from \ref{eq:eq5} (equation (5) of the main text) as 
follows: Equation \ref{eq:eq5} reads:
\begin{equation}
\dot Q_\alpha=\sum_{\beta} 
\int_0^\infty \omega d\omega \dot {\mathcal Q}_{\alpha\beta}(\omega)\coth(\omega/2k_BT_\beta).\nonumber
\end{equation}
Using the conservation equation $\sum_\beta \dot Q_{\alpha\beta}=0$ we find that 
\begin{equation}
\dot {\mathcal Q}_{\alpha\alpha}(\omega)=-\sum_{\beta\neq\alpha}\dot {\mathcal Q}_{\alpha\beta}(\omega). 
\end{equation}
Replacing this into \ref{eq:eq5}, and using the identity $\coth x=1+2n(x)$ we find
\begin{equation}
\dot Q_\alpha=-2\sum_{\beta\neq\alpha} 
\int_0^\infty \omega d\omega \dot {\mathcal Q}_{\alpha\beta}(\omega)(n_\alpha(\omega)-n_\beta(\omega)),\nonumber
\end{equation}
which is nothing but equation (7) of the main text.

\subsection{5) Generalized fluctuation-dissipation theorem}

Here we will prove the identity $2\omega\re\hat{\gamma}(i \omega) = {\pi}I(\omega)$, which, as mentioned in the main text, can be interpreted as a generalized version of the fluctuation-dissipation theorem. The Laplace transform of the damping kernel $\gamma(t)$ defined in \ref:{eq:gamma} is:
\begin{equation}
\label{eq:gamma-hat}
\hat{\gamma}(s) = \int_0^{\infty} dt \, \int_0^{+\infty} d\omega' \, \frac{I(\omega')}{\omega'} \cos(\omega' t) e^{-s t}. 
\end{equation}
Evaluating at $s = i \omega$ and taking the real part, we find:
\begin{equation}
\label{eq:gamma-hat}
\re\hat{\gamma}(i \omega) = \int_0^{\infty} dt \, \int_0^{\infty} d\omega'  \frac{I(\omega')}{\omega'} \cos(\omega' t) \cos(\omega t).
\end{equation}
As the integrand is an even function of $t$, it can be rewritten as:
\begin{equation}
\label{eq:gamma-hat}
\re\hat{\gamma}(i \omega) = \frac{1}{2} \int_0^{+\infty} d\omega' \, \frac{I(\omega')}{\omega'} \int_{-\infty}^{+\infty} dt \, \cos(\omega' t) \cos(\omega t),
\end{equation}
which can be transformed into:
\begin{align}
\label{eq:gamma-hat}
\re\hat{\gamma}(i \omega) &= \frac{\pi}{2} \int_0^{\infty} d\omega'  \frac{I(\omega')}{\omega'} (\delta(\omega' - \omega) + \delta(\omega' + \omega)). 
\end{align}
Thus, for positive $\omega$ we get:
\begin{equation}
\label{eq:supp-fd}
\operatorname{Re} \hat{\gamma}(i \omega) = \frac{\pi}{2 \omega} I(\omega),
\end{equation}
which is what we wanted to show.


\begin{thebibliography}{10}

\bibitem{SM} Supplementary material. 

\bibitem{Kosloff1} A. Levy and R. Kosloff, Phys. Rev. Lett. 108, 070604 (2012); A. Levy, R. Alicki and R. Kosloff "Quantum absorption refrigerators and the third law of thermodynamics", arXiv:1205.1347 (2012).

\bibitem{KimMahler} M. Michel, G. Mahler and J. Germer, {\em Phys. Rev. Lett.} \textbf{95}, 180602 (2005); K. Saito, {\em Europh. 
Lett.} \textbf{61}, 34 (2003). I. Kim and G. Mahler, 
{\em Phys. Rev.} \textbf{E81}, 1 (2010); J. Gemmer,
M. Michel and G. Mahler, \textit{Quantum Thermodynamics}.
(Springer, 2010).

\bibitem{Kuritzki} M. Kolar, D. Gelbwaser-Klimovsky,  R. Alicki and G. Kurizki, \textit{Quantum bath refrigeration towards absolute zero: unattainability principle
challenged}, arXiv:1208.1015.

\bibitem{others} R. Kosloff, J. Chem. Phys. {\bf 80}, 1625 (1984); S. Lloyd, Phys. Rev. {\bf A56}, 3374 (1997); R. Kosloff, E. Geva and J. M. Gordon, App. Phys. {\bf 87}, 8093 (2000).

\bibitem{HuPazZhang}
B.L. Hu, J.P. Paz and Y. Zhang, {\em Phys. Rev.} \textbf{D45},
2843, (1992) and references therein.

\bibitem{DavilaPaz}
L. D\'avila Romero and J.P. Paz, {\em Phys. Rev.} \textbf{A55}, 4070, (1997).

\bibitem{FlemingHuRoura}
G. Fleming, B.L. Hu and A. Roura, 
\newblock {\em Ann. Phys.} \textbf{100}, (2011).

\bibitem{ZurekPaz} J.P. Paz and W.H. Zurek;  \newblock {\em Fundamentals of Quantum Information}, Springer, 77-148 (2002). 

\bibitem{Dhar} B. Dhar, \newblock {\em Adv. In Phys.} \textbf{ 57}, 457 (2008).

\bibitem{Heffnergroup} T. Pruttivarasin, M. Ramm, I. Talukdar, A. Krouter and H. Haffner, \newblock {\em New Jour. Phys.} \textbf{13}, 075012  (2011).

\bibitem{LinDuan} G.D. Lin and L.M. Duan, 
\newblock {\em New Jour. Phys.} \textbf{13}, 075015  (2011).

\bibitem{biomodel} 
M. Mohseni, P. Rebentrost, S. Lloyd and A. Aspriu-Guzik, 
{\em Jour. Chem. Phys.} \textbf{129}, 174106 (2008); M.B. Plenio and S.F. Huelga, New J. Phys. {\bf 10}, 113019 (2008).

\bibitem{FV}
R.~P. Feynman and F.~L. Vernon, 
{\em Ann. Phys.} \textbf{24}, 118 (1963); see also H. Grabert, P. Schramm and G.L. Ingold; {\em Phys. Rep.} \textbf{168}, 115 (1988); A.O. Caldeira and A.J. Leggett, {\em Physica A} \textbf{121}, 567 (1983). 

\bibitem{Lili}
L. Arrachea, M. Moskalets and M. Martin--Moreno, 
\newblock {\em Phys. Rev} \textbf{B75}, 245420 (2007).

\end{thebibliography}
\end{document}